\documentclass{article}
\begin{document}
\baselineskip20pt

\begin{centering}
{\bf Relativistic Effects in the Motion of the Moon}\\
\vspace{1in}
Bahram Mashhoon \\ Department of Physics and Astronomy \\ University of
Missouri-Columbia \\ Columbia, Missouri  65211, USA \\
\vspace{.5in}
and\\
\vspace{.5in}
Dietmar S. Theiss \\ Institute for Theoretical Physics \\ University of
Cologne \\ 50923 K\"{o}ln, Germany \\
\vspace{1in}
{\bf Abstract}\\
\end{centering}
\vspace{.5in}
The main general relativistic effects in the motion of the Moon are briefly
reviewed.
The possibility of detection of the solar {\em gravitomagnetic}
contributions to the
mean motions of the lunar node and perigee is discussed. \\
\pagebreak
\newpage
\section{Introduction}
In a recent paper, Gutzwiller has provided an admirable
review of the oldest three-body problem, namely, the Sun-Earth-Moon
system [1].  Some work on the relativistic theory is mentioned
in his paper; however, in view of the recent advances in  relativistic
celestial mechanics this subject deserves a more complete discussion.
Here we provide a brief description of the main relativistic effects.

     The lunar laser ranging experiment has opened up the possibility
of measuring relativistic effects in the motion of the Moon; indeed,
the agreement between the standard general relativistic model that
contains over a hundred model parameters and the ranging data accumulated
over the
past three decades is excellent [2,3].  For instance, the post-fit
residuals in the Earth-Moon distance are at the centimeter level
[2,3].  Simple theoretical estimates lead to the conclusion that the
main relativistic effects in the lunar theory are due to the spin-orbit
coupling of
the Earth-Moon system in the gravitational field of the Sun.  The
post-Newtonian
influence of the solar field on the lunar motion consists of terms that can
be classified as either harmonic (i.e. periodic) or
secular (i.e. cumulative) in time.  It turns out to be very difficult in
practice to separate the harmonic terms from the corresponding Newtonian
terms with the same periodicities.  In effect, the existence of the
post-Newtonian
harmonic terms leads to small relativistic corrections in the numerical
values of certain model parameters
that are thereby adjusted by a fit to the ranging data.  To give an example of
such harmonic effects, we mention our prediction of a 6 cm relativistic tidal
variation in the Earth-Moon distance with a period of $1/2$ synodic month [4].

     The main secular terms turn out to be essentially due to the
precessional motion of the
Earth-Moon orbital angular momentum in the field of the Sun.  The
Earth-Moon system
can be thought of as an extended gyroscope in orbit
about the Sun; we are interested in the description of the motion of the
spin axis
of this gyroscope with respect to the ``fixed'' stars (i.e. the sidereal
frame).  An ideal
pointlike test gyroscope carried along a geodesic orbit would exhibit, in
the post-
Newtonian approximation, geodetic precession due to the orbital motion
around the mass
of the source as well as gravitomagnetic precession due to the intrinsic
rotation of the source;
however, the finite size of the gyroscope in this case (i.e. the orbit of
the Moon about the Earth) leads to additional
tidal effects.  The main post-Newtonian
gravitoelectric effect, i.e. geodetic precession, results in the advance of
the Moon's node and
perigee by about 2 arcseconds per century as first predicted by de Sitter
already
in 1916 [5].  This motion has been measured by Shapiro {\em et al.} with an
accuracy of about one percent [6].  It is a relativistic three-body effect;
therefore,
we consider in the next section the restricted three-body problem in
general relativity and briefly indicate,
in particular,  the more subtle
post-Newtonian gravitomagnetic contributions
to the motions of the Moon's node and perigee that are caused by the
rotation of the Sun; indeed, solar rotation induces
{\em cumulative} relativistic tidal effects in the Earth-Moon system [4].\\

\section{Restricted Three-Body Problem in General Relativity}
In our previous work [4], we developed a new scheme for the approximate
treatment of the
restricted three-body problem in general relativity.  This
coordinate-invariant approach is particularly
useful for a reliable theoretical description of relativistic (solar) tidal
effects in the motion of the Earth-Moon
system. We assume that the Moon follows a geodesic in the gravitational
field generated by the Earth and the Sun.
This field may be calculated as follows: we first imagine that the Earth
follows a geodesic in the solar field.
Along this geodesic, we set up a geocentric Fermi coordinate system.  This
system, which involves the tidal field of the Sun,
is then enhanced by taking due account of the field of the Earth in the
linear approximation.
Tidal effects in general relativity involve the projection of the Riemann
tensor
onto the tetrad frame of the measuring device.  Consider the tidal matrix
for a test
system (``Earth'') in free fall in the gravitational field of a rotating
mass (``Sun'').  In the standard
first-order post-Newtonian treatment, the spatial axes of the local tetrad
frame along the orbit are obtained by boosting
the background Minkowski axes and adjusting scales to maintain
orthonormality; the resulting
tidal matrix for an approximately circular geodesic orbit turns out to be
sinusoidal in time [7].
In this case, the  tetrad frame is not parallel-transported, but its motion
involves
the Lense-Thirring orbital precession as well as the geodetic (i.e. de
Sitter-Fokker) precession of the spatial
axes.  Once the parallel transport of the spatial axes along the orbit is
imposed, the gravitomagnetic (i.e. Schiff)
precession of the spatial axes would also appear in the first
post-Newtonian order.
In this order, the tidal matrix for the parallel-transported axes contains
a secular term as well
that must therefore be a direct consequence of the Schiff precession of the
spatial axes [8], in agreement
with our previous work [9-11].  The linear growth of this gravitomagnetic
contribution to the tidal
field poses a problem for the first post-Newtonian approximation:  the
non-Newtonian ``off-diagonal''
part of the tidal matrix can diverge in time [9-11].  To avoid this
limitation, we have developed a post-Schwarzschild
treatment of  gravitomagnetic tidal effects; indeed,  the concept of
relativistic nutation provides a natural resolution of this difficulty by
limiting the
temporal extent of validity of the post-Newtonian approximation [9-11].

     Imagine, for instance, a set of three orthogonal test gyroscopes
falling freely along an
inclined circular geodesic orbit with constant radius $r$ (``astronomical
unit'') about a slowly
rotating central body (``Sun'') with mass $M$ and proper angular momentum
$J$.  The motion
of the spin axes of these torque-free gyroscopes, which constitute a local
inertial frame (i.e.
the geocentric Fermi frame), is essentially governed by the equations of
parallel transport along the geodesic
orbit.  By solving these equations using the post-Schwarzschild
approximation scheme that takes $M$ into account to all orders, it can be
shown that the average motion of the gyroscope axes with respect to an
effective Newtonian (i.e. sidereal) frame
consists of a gravitoelectric precessional motion---i.e. geodetic
precession that was first completely analyzed by Fokker---
together with a complex gravitomagnetic motion that can be loosely
described as a combination
of precessional movement and a harmonic nodding  movement.  The latter
motion is a new relativistic effect of a rotating mass and has been
referred to as {\em relativistic nutation} [11].
In the post-Newtonian approximation, the nutational terms over a limited
time combine with the
other gravitomagnetic precessional terms to give the Schiff precession.  To
see how this comes
about, let us denote by $\tau$ the proper time of the geodesic orbit and
consider a vector
normal to the orbital plane (ecliptic) at the beginning of measurement $(\tau = 0)$.  Relativisticnutation is a periodic variation of the angle between this
vector and a gyroscope axis that is
Fermi propagated along the orbit.  The leading contribution of relativistic
nutation
to this angle can be written as
\begin{equation}
\Theta_{n}\approx\xi[\sin(\eta_{0}+\omega_{F}\tau)-\sin \eta_{0}]\sin\alpha   ,
\end{equation}
where $\eta_{0}$ is the azimuthal position of the Earth in the ecliptic at
$\tau=0$ measured from
the line of the ascending nodes and $\xi=J/Mr^{2}\omega$.  Here $\omega,
\omega^{2}=GM/r^{3}$,
approximately describes the orbital frequency in the absence of rotation
and $\alpha$ denotes the
inclination of the orbit with respect to the equatorial plane of the Sun
[12].  The frequency of this
nutational oscillation is the Fokker frequency
$\omega_{F}\approx\frac{3}{2}\epsilon  \omega$, where $\epsilon
=GM/c^{2}r$.  The nutation amplitude, $\xi\sin\alpha$, does not depend on
the speed of light $c$.
This remarkable fact can be traced back to the occurrence of a small
divisor [9-11] involving the Fokker
frequency.  In the post-Newtonian limit of the post-Schwarzschild
approximation, Eq. (1) reduces to
$\Theta_{n}\sim\omega_{n}\tau$, which represents a {\em precessional}
motion with frequency $\omega_{n}=
\xi \omega_{F} \sin \alpha \cos \eta_{0}$ about a direction opposite to
that of orbital velocity at $\tau = 0$.
Thus, relativistic nutation reduces to a {\em part} of the Schiff
precession in the first post-Newtonian approximation.  It
follows from this analysis that the first post-Newtonian approximation
breaks down over timescales of the order of Fokker
period $\tau_{F}=2\pi/\omega_{F}$; however, this fact does not diminish
the usefulness of the first post-Newtonian approximation for the
description of observations in
the solar system since in this case the Fokker period is almost
immeasurably long
(e.g. $\tau_{F}\simeq$ 67 million years for the motion of the Earth about
the Sun).

     Let us consider the influence of the gravitomagnetic field of the
central body (``Sun'') on
the relative acceleration of two nearby test particles (``Earth'' and
``Moon'') moving along the circular geodesic
orbit.  The dominant contributions of the gravitomagnetic field of the
central body to the tidal
matrix, first calculated by the authors [9-11], are proportional to
\begin{equation}
\omega^{2}\xi\sin\alpha\sin\left(\frac{1}{2} \omega_{F}  \tau\right) ,
\end{equation}
which is directly proportional to the amplitude of relativistic nutation
($\xi\sin\alpha$) and
exhibits a maximum (at $\tau=\tau_{F}/2$) that is independent of the speed
of light $c$.  It follows
from Eq. (2) that to first order in $\omega_{F}  \tau\ll1$, the dominant
gravitomagnetic amplitude
varies linearly with $\tau$.  This secular amplitude originates from a
coupling of the {\em nutation
part} of Schiff precession  with the amplitude ($\sim\omega^{2}$) of the
Newtonian contribution to the
gravity gradient [13].  It should be mentioned in passing that the
relativistic quadrupole
contributions to the tidal matrix have properties quite similar to the
gravitomagnetic tidal effect
described here [10].

     Let us now turn to the potentially observable effects of the solar
gravitomagnetic field on the lunar
motion. The lunar path is determined by the Newton-Jacobi equation
\begin{equation}
\frac{d^{2}x^{i}}{d\tau^{2}}+\frac{Gm}{R^{3}}x^{i}=-K_{ij}(\tau)  x^{j},
\end{equation}
where $x^{i},i=1,2,3,$ represent the geocentric Fermi coordinates of the
Moon, $m$ is the total mass of
the Earth-Moon system and $R(\tau)$ denotes the Earth-Moon distance
depending on the proper
time $\tau$ measured along the geocentric path around the Sun.  Here $K$ is
the tidal matrix.  Equation (3) describes the
motion of the Moon with respect to a geocentric local inertial frame [14].
Using the equation
of relative motion (3), we have calculated---among other things---the
influence of the tidal field
of the Sun on the orbital angular momentum of the Moon with respect to the
Earth.  To
express the result with respect to the sidereal frame, we choose as our sidereal reference framethe geocentric Fermi frame at $\tau=0$.  This frame is
related to the Fermi frame at time $\tau$ by
a rotation matrix that incorporates the relativistic precession and
nutation of the Fermi
frame with respect to the sidereal frame.  In the first post-Newtonian
approximation,
this motion reduces to a (Fokker plus Schiff) precession.  Let $D$ denote
this rotation matrix, then
${\cal L}_{i}=D_{ij}L_{j}$,
where the sidereal components $({\cal L}_{i})$ of the orbital angular
momentum are obtained from a
transformation of the geocentric components $(L_{j})$ with
\begin{equation}
D_{ij}=\delta_{ij}-\epsilon_{ijk}\Phi_{k}\hspace{.1in},\hspace{.1in} {\bf
\Phi}=\int^{\tau}_{0}\omega_{FS}(\tau^{\prime})d\tau^{\prime}.
\end{equation}
Here the analysis is limited to the first post-Newtonian approximation and
$\omega_{FS}$ represents the
frequency of (Fokker plus Schiff) precession.  The direction of Schiff
precession is not fixed along the
Earth's orbit; therefore, ${\bf \Phi}$ contains (cumulative) secular terms
(which represent simple precession)
together with (harmonic) nutational terms of frequency $2\omega$ and
amplitude of order $\alpha\epsilon\xi$.
Averaging over the latter terms, the dominant secular terms in ${\bf \Phi}$
are given by
\begin{equation}
\Phi_{1}\sim\frac{1}{2}\alpha\epsilon\xi\omega\tau\sin\eta_{0}\hspace{.1in}
,\hspace{.1in}
  \Phi_{2}\sim\left(-\frac{3}{2}+\xi\right)\epsilon\omega\tau\hspace{.1in}
,\hspace{.1in}
 \Phi_{3}\sim\frac{1}{2}\alpha\epsilon\xi\omega\tau\cos\eta_{0}\hspace{.1in}
\end{equation}
with respect to the geocentric Fermi frame at $\tau=0$, which has its first
axis essentially along the radial position of the
Earth, the third axis approximately along the direction of motion of the
Earth and the second axis normal to the ecliptic
(in a direction opposite to the Earth's orbital angular momentum about the
Sun).  We note that for $\xi=0$, Eq. (5) expresses
the de Sitter-Fokker effect that has been observed by Shapiro {\em et al}.
[6].  To illustrate our approach, let us use Eq. (3)
to determine the value of {\bf L}, which is the angular momentum of the
Moon in a circular orbit about the Earth with respect to
the Fermi frame, averaged over orbital motions of the Earth about the Sun
(with frequency $\omega$) and the Moon about
the Earth (with frequency $\Omega$).  Then
\begin{equation}
\frac{d\langle{\bf L}\rangle}{d\tau}=\tilde{\omega}\times{\bf
L_{0}}\hspace{.1in},
\end{equation}
where ${\bf L_{0}}$ is the unperturbed orbital angular momentum with
respect to the geocentric Fermi frame and $\tilde{\omega}$
is given by
\begin{equation}
\tilde{\omega}^{1}\approx-\tilde{\omega}_{0}\alpha\epsilon\xi\left(2\sin\eta_{0}
+\frac{3}{2}\omega\tau\cos\eta_{0}\right),
\end{equation}
\begin{equation}
\tilde{\omega}^{2}\approx\tilde{\omega}_{0}(1-6\epsilon\xi),
\end{equation}
\begin{equation}
\tilde{\omega}^{3}\approx-\tilde{\omega}_{0}\alpha\epsilon\xi\left(2\cos\eta_{0}
-\frac{3}{2}\omega\tau\sin\eta_{0}\right)
\end{equation}
to first order in the tidal perturbation characterized by the Newtonian
regression frequency $\tilde{\omega}_{0}=3\omega^{2}/4\Omega$,
which corresponds to a period of nearly 18 years [15,16].  It is clear from
Eqs. (4)-(9) that the motion of $\langle{\cal L}\rangle$
can be expressed as a {\em Newtonian} regression modulated by long-term
(secular) {\em relativistic} perturbations characterized
by the de Sitter-Fokker, Schiff and gravitomagnetic tidal effects.  To
illustrate this point, let us assume for the sake of
simplicity that in the absence of relativistic effects the lunar orbital
angular momentum undergoes a steady regression of
frequency $\tilde{\omega}_{0}$ and that once relativistic effects are
included the average motion in the Fermi frame is one of
precession with the frequency given by Eqs. (7)-(9).  It then follows that
the expression for $\langle{\cal L}_{2}\rangle$, i.e.
the average of the second sidereal component of the lunar orbital angular
momentum, contains a dominant gravitomagnetic contribution
of the form
\begin{equation}
\langle{\cal L}_{2}\rangle_{\rm secular}\approx2\alpha\beta\epsilon\xi(\mu
R^{2}_{0}\Omega)\omega\tau\sin(\eta_{0}+\zeta_{0}
-\tilde{\omega}_{0}\tau),
\end{equation}
where $\beta, \mu, R_{0}$ and $\zeta_{0}$ denote, respectively, the
inclination of the lunar orbit with respect to the
ecliptic  $(\approx 5^{\circ})$, the mass of the Moon, the mean Earth-Moon
separation and the longitude of the ascending node of
the orbit of the Moon measured from the first axis of the sidereal frame.
These simple considerations that are based on an initial
circular orbit only indicate the nature of the secular terms involved;
clearly, extensive calculations are necessary for a
complete treatment.

\section{Discussion}
The results of our theoretical work are of particular interest for the
description of dominant relativistic gravitational effects
in the motion of the Moon, especially the gravitomagnetic tidal component
of the motion of the orbital angular momentum of the
Moon.  It is important to point out that the eccentricities of the orbit of
the Earth around the Sun and the orbit
of the Moon about the Earth should be taken into account; we have ignored
them in our preliminary analysis [16].
As lunar laser ranging data further accumulate, it may become possible in
the future to deduce the angular momentum of the
Sun from the measurement of the solar gravitomagnetic contributions to the
mean motions of the lunar node and perigee.

It is interesting to compare our secular gravitomagnetic tidal terms with
hypothetical terms that might indicate a temporal
variation of the gravitational ``constant'' $G$.  Our results have thus far
been based on a secular term proportional to
$\tau$ in the tidal matrix $K$ in Eq. (3); however, as can be seen from the
middle term in Eq. (3), similar effects could be
produced if such a term appears in $G$ instead.  We have shown that our
predictions are similar to a variation of $G$ in Eq. (3)
at the level of $10^{-16}$ yr$^{-1}$; moreover, there are significant
differences between the two effects that can be used to
separate them [4,9].  The present upper limit on $|\dot{G}/G|$ is at the
level of $10^{-12}$ yr$^{-1}$; therefore, it may be a long
while before the gravitomagnetic effects in the motion of the Moon become
detectable.\\

\noindent{\bf Acknowledgments\\}
We would like to thank Friedrich Hehl for fruitful discussions. We are
also grateful to Sergei Kopeikin for helpful
comments.  This work was supported in part by the Deutsche
Forschungsgemeinschaft, Bonn.


\newpage
\begin{thebibliography}{99}
\bibitem{[1]}  M.C. Gutzwiller, Rev. Mod. Phys. {\bf 70}, 589 (1998).
\bibitem{[2]}  J.O. Dickey {\em et al.}, Science {\bf 265}, 482 (1994).
\bibitem{[3]}  K. Nordtvedt, these proceedings; Phys. Today {\bf 49}, no. 5, 26 (1996).
\bibitem{[4]}  B. Mashhoon and D.S. Theiss, Nuovo Cimento B {\bf 106}, 545
(1991).  A misprint in Eq. (64) of this paper
should be corrected:  The overall sign of the right-hand side of Eq. (64)
should be negative.  This correction would make
that equation compatible with Eq. (8) of the present paper.
\bibitem{[5]}  W. de Sitter, Mon. Not. Roy. Astron. Soc. {\bf 77}, 155 (1916).
\bibitem{[6]}  I.I. Shapiro, R.D. Reasenberg, J.F. Chandler and R.W.
Babcock, Phys. Rev. Lett. {\bf 61}, 2643 (1988).  See
also B. Bertotti, I. Ciufolini and P.L. Bender, Phys. Rev. Lett. {\bf 58},
1062 (1987).
\bibitem{[7]}  B. Mashhoon, H.J. Paik and C.M. Will, Phys. Rev. D {\bf 39},
2825 (1989).
\bibitem{[8]}  C.A. Blockley and G.E. Stedman, Phys. Lett. A {\bf 147}, 161
(1990).
\bibitem{[9]}  B. Mashhoon and D.S. Theiss, Phys. Rev. Lett. {\bf 49}, 1542
(1982); Phys. Lett. A {\bf 115}, 333 (1986); B.
Mashhoon, Gen. Rel. Grav. {\bf 16}, 311 (1984).
\bibitem{[10]} D.S. Theiss, Ph.D. thesis, University of Cologne (K\"{o}ln,
1984); Phys. Lett. A {\bf 109}, 19, 23 (1985); in
{\em Relativistic Gravity Research}, edited by J. Ehlers and G. Sch\"{a}fer
(Springer, Berlin, 1992), p. 131; in {\em Proc.
First William Fairbank Meeting on Relativistic Gravitational Experiments in
Space}, edited by M. Demianski and C.W.F. Everitt
(World Scientific, Singapore, 1993), p. 227; in {\em Proc. Seventh Marcel
Grossmann Meeting on General Relativity}, edited by
R.T. Jantzen and G. Mac Keiser (World Scientific, Singapore, 1996), p.
1555, p. 1558.
\bibitem{[11]} B. Mashhoon, Found. Phys. {\bf 15} (Bergmann Festschrift),
497 (1985).  In this paper a misprint in Eq. (22) must
be corrected: $r_{0}$ in the denominator of the last term must be replaced
by $c$.  Furthermore, in Eqs. (11) and (13) the
temporal components must be divided by $c$.
\bibitem{[12]} For the motion of the Earth about the Sun,
$\alpha\approx7^{\circ}$, $\epsilon\approx10^{-8}$ and $\xi\approx2
\times10^{-5}$ based on the standard value of solar angular momentum [cf.
C.W. Allen, {\em Astrophysical Quantities}, 3rd ed.
(Athlone, London, 1973)].  The Fermi frame adopted in this paper is such
that the nutation vanishes at $\tau=0$.  This frame
can be obtained from the results given in Ref. [11]; see, especially, p. 506.
\bibitem{[13]} In this connection, it is also important to note that
$\omega_{n}$ vanishes for an equatorial orbit $(\alpha=0)$
in contrast to the frequency of the {\em full} Schiff precession.
\bibitem{[14]} The precise definition of the notion of a local geocentric
frame was formulated in [11] and further developed in
S.M. Kopeikin, Celestial Mech. {\bf 44}, 87 (1988) and V.A. Brumberg and
S.M. Kopeikin, Nuovo Cimento B {\bf 103}, 63 (1989).
\bibitem{[15]} It follows from a more complete treatment of the Newtonian
problem that the mean motion of the lunar node can be
characterized by a backward movement of frequency $\tilde{\omega}_{0}N$,
which corresponds to a period of about 18.61 years.
Similarly, the mean motion of the perigee can be characterized by a forward
movement of frequency $\tilde{\omega}_{0}P$, which
corresponds to a period of about 8.85 years.  The theoretical expressions
for $N$ and $P$ are rather complicated and depend on
$\omega/\Omega$ as well as the orbital eccentricities, etc.  The first two
terms of $N$ and $P$ in terms of $\nu=\omega/\Omega$
are given by $N=1-3\nu/8-\cdots$ and $P=1+75\nu/8+\cdots$.  A detailed
discussion of this subtle problem is given by D. Brouwer
and G.M. Clemence, {\em Celestial Mechanics} (Academic Press, New York,
1961), Ch. 12, especially pp. 320-323.
\bibitem{[16]} If our analysis is extended by including a slight
eccentricity for the lunar orbit, then the motion of the
Runge-Lenz vector indicates an average forward precession of perigee with
frequency $\tilde{\omega}_{0}(1-6\epsilon\xi)$.  The
relativistic gravitoelectric correction to this result is of order
$\epsilon^{2}$ in the post-Newtonian scheme; however, if the
standard Schwarzschild coordinates are used in the analysis (cf. Ref. [4]),
$\tilde{\omega}_{0}\rightarrow\tilde{\omega}_{0}(1+3\epsilon)$,
so that a slight increase in the forward (backward) precession rate of the
perigee (node) would occur to first order in the
tidal perturbation.  This supersedes (and partly corrects) an assertion in
Ref. (21) of our Nuovo Cimento paper (cf. Ref. [4])
regarding the comparison of our results with Robertson's ``solar effect.''
However, a more complete analysis of the relativistic
three-body problem is necessary for a reliable calculation of the motion of
the lunar perigee.
\end{thebibliography}
\end{document}